\documentclass[conference]{IEEEtran}
\usepackage{times}

\usepackage[numbers]{natbib}
\usepackage{multicol}
\usepackage[bookmarks=true]{hyperref}
\usepackage{amsmath}
\usepackage{amssymb}
\usepackage{graphicx}
\usepackage[dvipsnames]{xcolor}
\usepackage{lipsum}

\begin{document}

\title{Transformer Based Time-Series\\ Forecasting For Stock}


\author{

\authorblockN{Shuozhe Li\footnotemark *}
\thanks{These two authors contributed equally}
\authorblockA{UT Austin\\
Email: shuozhe.li@utexas.edu}

\and

\authorblockN{Zachery B Schulwolf\footnotemark *}
\thanks{These two authors contributed equally}
\authorblockA{UT Austin\\
zachery@utexas.edu}

\and

\authorblockN{Risto Miikkulainen}
\thanks{These two authors contributed equally}
\authorblockA{UT Austin\\
risto@cs.utexas.edu}

}



%

\maketitle

\begin{abstract}
To the naked eye, stock prices are considered chaotic, dynamic, and unpredictable. Indeed, it is one of the most difficult forecasting tasks that hundreds of millions of retail traders and professional traders around the world try to do every second even before the market opens. With recent advances in the development of machine learning and the amount of data the market generated over years, applying machine learning techniques such as deep learning neural networks is unavoidable. In this work, we modeled the task as a multivariate forecasting problem, instead of a naive autoregression problem. The multivariate analysis is done using the attention mechanism via applying a mutated version of the Transformer, "Stockformer", which we created.
\end{abstract}

\IEEEpeerreviewmaketitle

\renewcommand{\thefootnote}{\fnsymbol{footnote}}
\footnotetext[1]{These authors contributed equally to this work.}

\section{Introduction}
Predicting the financial time series such as stock price means predicting the behavior of the stock price steps ahead of the series with the help of various variables. By knowing the behavior of the stock price ahead, one can take the advantage of it to beat the market. Thereby, the benefit of beating the market attracts the creation of numerous methods to predict the stock price. But, in the view of traditional finance, according to the Efficient Market Hypothesis, the current stock prices only reflect the current market information, and unless by knowing all the new market information ahead, it is impossible to predict the new prices. This implies that the stock cannot be accurately predicted using historical values. However, researches like \cite{brock1992simple} find techniques such as trading-range breaks and moving averages proves that prices can be predicted to a certain degree. Hence, there has not been a conclusion drawn on the predictability of stock price. In addition, with the emergence of artificial neural network, evidence suggested that time series forecasting models \cite{KOHZADI1996169} are suitable for the price prediction task. Meanwhile, well known work such as \cite{barr1998persuasive} has proved a considerable level of market inefficiency is present in a wide range of markets. With the occurrence of market inefficiency, it is reasonable to assume there are relationships among the stock prices of companies within one type of industry. By assuming the inefficiency of the market and taking the advantage of the relationships among the stock prices, the trader can make advantageous decisions to beat the market. To find the relationships among the prices across the time and predict the stock price of the target company, this work introduces a Transformer based multivariate to one time series forecasting model "Stockformer"\cite{zhou2021informer}.

\section{Related Work}
In the view of traditional finance, there are two classes of approaches to predict stock market. They are technical analysis and fundamental analysis \cite{as2013study}. The technical analysis assumes the market value of a stock is solely determined by the interaction of supply and demand factors operating in the market. Technical analysis thinks that the market actions which decide supply and demand factors tends to repeat themselves according to history. Fundamental analysis studies the macroeconomic data that can affect the the stock price. It focuses on the factors including overall economic and industry conditions and finical statement of the company. 
    
Financial data like stock are not generally described by simple linear structure for random walks or noise. Neural network, in theory, when compare to conventional statistical method, is more robust to inaccurate and missing data, and, according to Universal Approximation Theorem, neural network is able to approximate any complex nonlinear pattern from the data. Therefore, it has been an active research area to try to use neural networks to predict financial market. In the early years, simple multi-layer perceptron and probabilistic neural network \cite{501826} were created to perform predictions. Meanwhile, researchers were also try to combine the classical fundamental and technical analysis with multi-layer perceptron \cite{8488440} which create a hybrid model that outperform the results obtained from the technical and fundamental analysis in isolation. According to the paper \cite{8488440}, this result also provides strong evidence to conclude that the market is not perfectly efficient. However, as \cite{ismail2019deep} points out, the issue with multi-layer perceptron is that the features learned are not time-invariant and the temporal information is lost.
    
To attack the above issues, convolutional neural network (CNN) plays an important role. Although CNNs are traditionally used for image and pattern recognition by extracting features from 2D data\cite{726791}, 1D CNNs can also learn spatially invariant features from the raw input time series\cite{wang2017time}. In addition, convolutional neural network can also be used for automatic feature extraction to capture the correlation which possibly exist between the stock market as well as other source of information such as technical indicators\cite{HOSEINZADE2019273}. On the other hand, recurrent neural network models (RNNs), including the two important variants, gated recurrent unit (GRU) \cite{chung2014empirical} and long short term memory (LSTM) \cite{hochreiter1997long} are designed to better process with the temporal (or sequential) information. When training the RNNs, the input signals pass through recurrent connections which memorizes the important features, and when it is deployed, the information in the memory can be used to forecast the future value\cite{gjylapirecurrent}. Nonetheless, RNNs are not good at extracting useful features from the input of each time stamp. As a result, researchers tries to combined the CNNs and the LSTM\cite{lu2020cnn}. Together, the combined CNN-LSTM network, the CNN is used to extract helpful features from intentionally selected data that relates to the stock. Then, the LSTM predicts the stock price with the extracted features\cite{lu2020cnn}. Although the accuracy has been improved significantly, there are several problems with this method. First, computing power has become a major bottleneck for deep learning. Researchers have always wanted to take advantage of parallel computing. However, the structure of the RNN is not appropriate for parallelization. Second, the information passed down from the early recurrent node is very likely to be forgotten if the input time sequence is very long. Therefore, to find a relationship between two time steps that are far from each other is very hard for RNNs structure.
\begin{figure}[h!]
    \centering
    \includegraphics[scale=0.5, width=50mm]{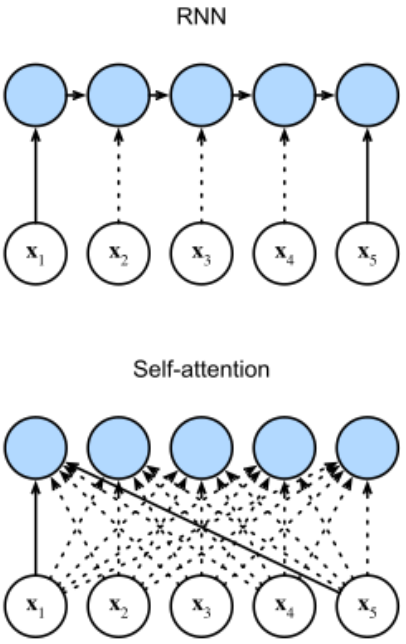}
    \caption{Finding dependencies on two positions in RNN requires traversing through recurrent unit;}
    \label{fig:RNN_Attention}
\end{figure}
\\
\noindent Aiming to solve the problems that the recurrent structure faces above, at 2017, researchers from Google created a novel architecture solely based on attention mechanism called \textit{Transformer} \cite{vaswani2017attention}. Consider having an input length of \textit{n}, when learning a long-range dependencies between two positions, the shorter the path forward and backward between any combination of positions for a signal to traverse, the easier it is to learn the dependencies \cite{hochreiter2001gradient}. As shown in Fig \ref{fig:RNN_Attention}, for RNN, it would take \textit{O(n)} to learn the dependencies because the signal has to traverse through all the recurrent unit between the two positions \cite{hochreiter2001gradient}. However, a self-attention layer only requires a constant number of executed operations\cite{vaswani2017attention}. This significantly decreased the difficulty for the network to learn the long-range dependencies. In addition, \textit{Transformer} architecture is highly parallelizable. During the calculation of scaled dot-product attention, the repetitive calculations can be removed by turning everything into huge matrix multiplications. Via utilizing GPU, this process can be accelerated parallelly. On the other hand, for RNN, it has to wait until the previous recurrent unit to finish its calculation.

In this project, we implemented \textit{Stockformer} on the top of the Transformer, discussed issues of naive Transformer, and changed the original architecture to fit with the financial ticker forecasting task.

\section{Problem Formulation}
Although the goal for the neural network is to predict the stock price ahead, the task for this project is to assist traders to make a profit in the end. Since we are assisting the human trader and not doing the high-frequency trading, we need to give the human trader enough time to react with the output of the model. Therefore, together with the need to capture enough variations at each time point for the neural network, we decide to use one hour as the time window for the model to make the prediction for the stock price. Hence, for every hour during the regular market hours, the model will output its price prediction. Based on the model's prediction, the trader can decide to buy or short the stock.

\begin{figure}[h!]
    \centering
    \includegraphics[width=\linewidth]{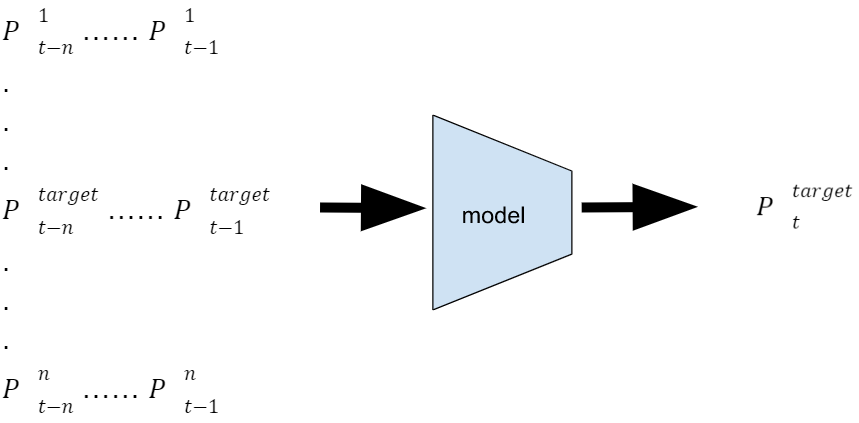}
    \caption{The inputs here are the price information which can be percents of change or real price value; The output can be the price information or just trend;}
    \label{fig:p1}
\end{figure}

\noindent To predict the stock price \textit{$p_{t}$} one hour ahead, the model will take in stock prices from \textit{$t-n$} to \textit{$t-1$} of several highly correlated stocks \textit{$p_{}^{0...n}$} and financial securities including the target stocks. This defines the task as a multivariate time series forecasting problem. As a special case, we are only forecasting one financial ticker or stock.

\section{Approach}
\subsection{Data Collection}
In the era of big data, people always say, "more data beats clever algorithms, but better data beats more". To have data that covers enough variations in the stock market for the network to learn the pattern while trying not to spend money on it is not easy as it seems. We tried several financial information platforms:
\subsubsection{\textit{Yahoo}}
\textit{Yahoo} only provides daily, weekly, and monthly data for all the financial securities. But, this is not fine-grained enough for our models to assist a human trader during the opening market hours. In addition, with only daily data for 10 years, there are only 3650 time stamp data points to train the Transformer structured neural network which is insufficient.
\subsubsection{\textit{alphavantage.co}}
Alpha Vantage is a very popular up-in-coming API provider for financial market data coming out of Y-Combinator. However, they only allow users to access 2 years of hourly data in the past. 
\subsubsection{\textit{alpaca.markets}}
Alpaca is another extremely popular trading platform that provides both data and trading APIs for retail automated traders. Their historical data API goes back 5 years but the prices are not adjusted (do not account for stock splits) and come from various providers (lesser quality).
\subsubsection{\textit{polygon.io}}
Polygon.io has powerful APIs that are able to provide information about the market status, news related to the stock, financial information for fundamental analysis, and even the options contract of stocks on the market. This gives huge potential for future development on our project. Moreover, it provides hourly data on the past 10 years at a good price.

\subsection{Data Preprocessing}
Although online platforms can provide enough amount of data to train the neural network, the data content and format are not perfect. Hence, data preprocessing is needed to clean the data and scale the data to a common range. In our case, different stocks and financial securities have different price ranges. Feeding data with different ranges directly into the neural network will cause the neurons difficult to learn. Therefore, we decided to change the value from the price to the percent of the closing price over the opening price of the hour. In addition, to stabilize the variance of the data and obtain the smoothed percent of change. A natural log transform is applied to the percentage.
\begin{equation}\label{Data_Preprocessing}
    Log Percent Change = ln(\frac{Closing Price}{Opening Price} + 1)
\end{equation}
\begin{equation}\label{Data_Preprocessing}
    Percent Change = \frac{Closing Price}{Opening Price}
\end{equation}

\subsection{Financial Securities With Causality}
As we mentioned in the introduction section, this project is built on the idea of assuming relationships of stocks in one industry. Taking the oil industry as an example, similar trends and patterns across different stocks can be discovered visually.

\begin{figure}[h!]
    \centering
    \includegraphics[width=\linewidth]{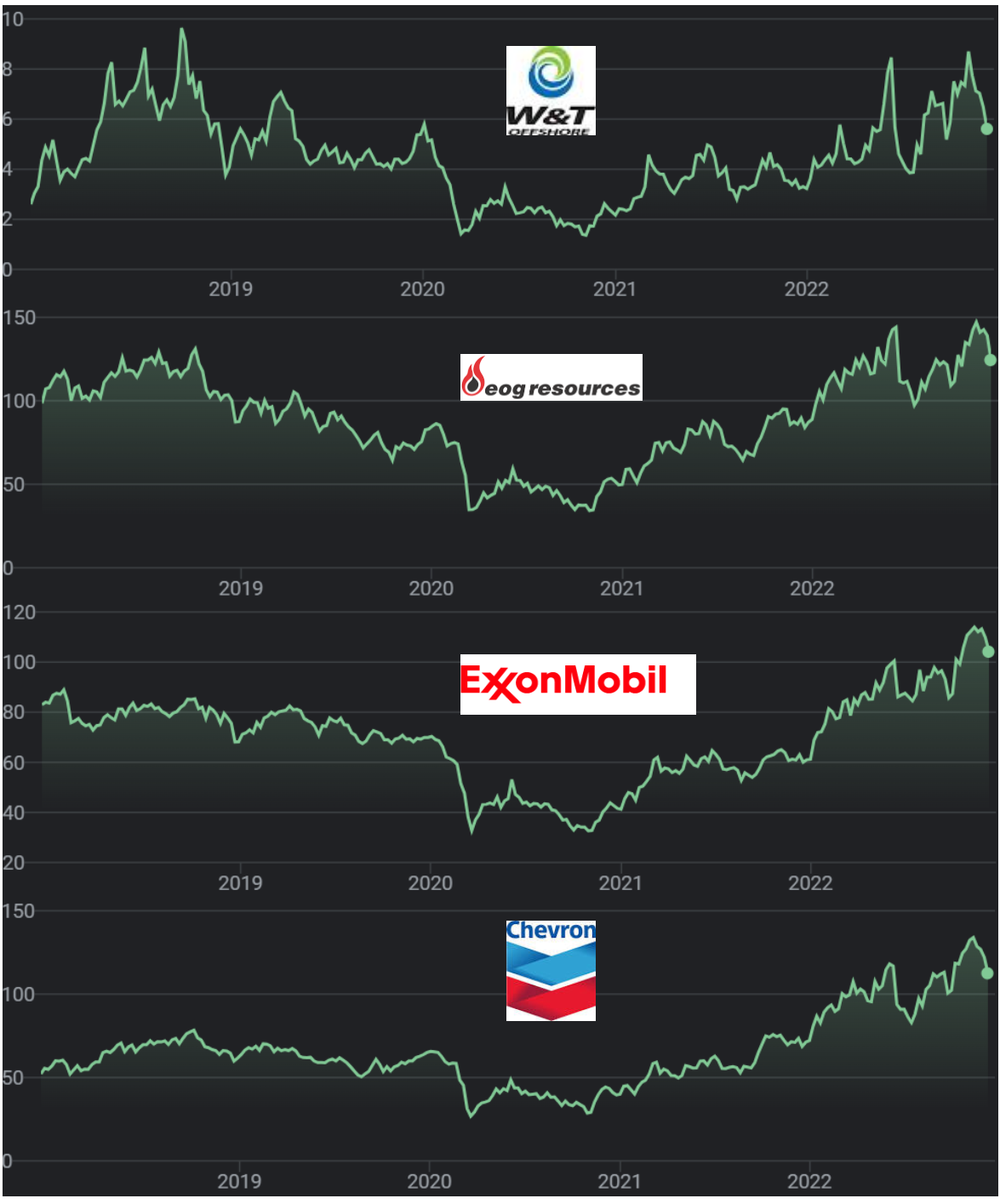}
    \caption{The stock prices of four American multinational oil and gas corporations in the past 5 years;}
    \label{fig:p2}
\end{figure}

\noindent From Fig \ref{fig:p2}, one can easily find that the overall trend among the stocks. However, in order to take advantage of this trend. We need to use this pattern within a smaller time window. In our case, we ask the question, if the stock price of ExxonMobil dropped 5 percent at 2 P.M., does this mean the stock price of Chevron will also drop some percent at 3 P.M.? In conventional statistical theory, this type of causality can be tested via Granger causality test \cite{diks2006new}. This test is based on the idea that if one time series is truly useful for forecasting another time series, then a statistical model that includes the past values of the first time series should be able to make more accurate predictions than a model that only uses the past values of the second time series.
\begin{table}[!ht]
    \centering
    \begin{tabular}{|l|l|l|l|l|l|l|l|}
    \hline
        ~ & XOM & CVX & COP & BP & PBR & WTI & EOG \\ \hline
        XOM\_Y & 1.0 & 0.3204 & 0.1484 & 0.5337 & 0.9651 & 0.5131 & 0.4394 \\ \hline
        CVX\_Y & 0.5223 & 1.0 & 0.0655 & 0.6965 & 0.2068 & 0.755 & 0.2261 \\ \hline
        COP\_Y & 0.1156 & 0.3724 & 1.0 & 0.4059 & 0.9479 & 0.109 & 0.126 \\ \hline
        BP\_Y & 0.0004 & 0.3159 & 0.0027 & 1.0 & 0.4099 & 0.5154 & 0.0044 \\ \hline
        PBR\_Y & 0.1228 & 0.6649 & 0.4954 & 0.4999 & 1.0 & 0.0096 & 0.1365 \\ \hline
        WTI\_Y & 0.0097 & 0.5562 & 0.094 & 0.4936 & 0.042 & 1.0 & 0.0211 \\ \hline
        EOG\_Y & 0.525 & 0.1245 & 0.3163 & 0.2442 & 0.586 & 0.0819 & 1.0 \\ \hline
    \end{tabular}
\caption{p-value of Granger Causality test when the lag is set to 4; p-Values lesser than the significance level (0.05) implies patterns in one stock price are approximately repeated by others (Y series) after or before some time; Y means the stock being forecasted;}
\end{table}

\noindent By summing and comparing the p-value on each row, predicting the stock price of W\&T Offshore is found to be most beneficial when the stock prices of other companies are in the model. Therefore, we choose to predict the stock price of W\&T Offshore and use neural network model to take advantage of the causality among the stock prices.

\section{Architecture}
When it comes to design a neural network based on Transformer architecture, there are many choices such as different ways of doing embeddings, encoders, and decoders. The following section of the report will focus on discussing the design choices for stockformer.
\subsection{Token Embedding Design}
In most of the use cases, the token embedding layer in a Transformer-based model learns a fixed-length vector representation of a variable-length sequence input. The Embedding layer will keep the sequence length while extracting more features from the input in each time step. During our development, we have two options for token embedding design.
\subsubsection{Fully Connected Based}
The embedding of the input sequence will be learned via several linear layers. However, the temporal information will be lost during the operation because, in order to keep the sequence length, the linear layers will be only learning the patterns among the financial securities in each time step and the relations related to time will be ignored.
\subsubsection{1D-CNN Based}
Assuming there will be \textit{$s_{in}^{}$} number of financial securities and data of \textit{n} time steps are known before the prediction. The 1D-CNN will have \textit{$s_{in}^{}$} channels as its input and \textit{$s_{out}^{}$} channels in its output. Hence, during the 1D convolutional operation, there will be separated kernels for each financial securities. With each financial securities, a kernel window will be sliding through the time steps to learn the temporal information in the sequence. In the end, the output channels will store the fine-grained temporal information learned from each financial securities. In addition, to keep sequence length, the 1D CNN layer has a kernel of 3, stride of 1, padding of 1. Meanwhile, with the padding mode set to circular. the edges of the data are "stitched" together to avoid boundary effects and can improve the accuracy of the convolutional layer.

\begin{figure}[h!]
    \centering
    \includegraphics[width=\linewidth]{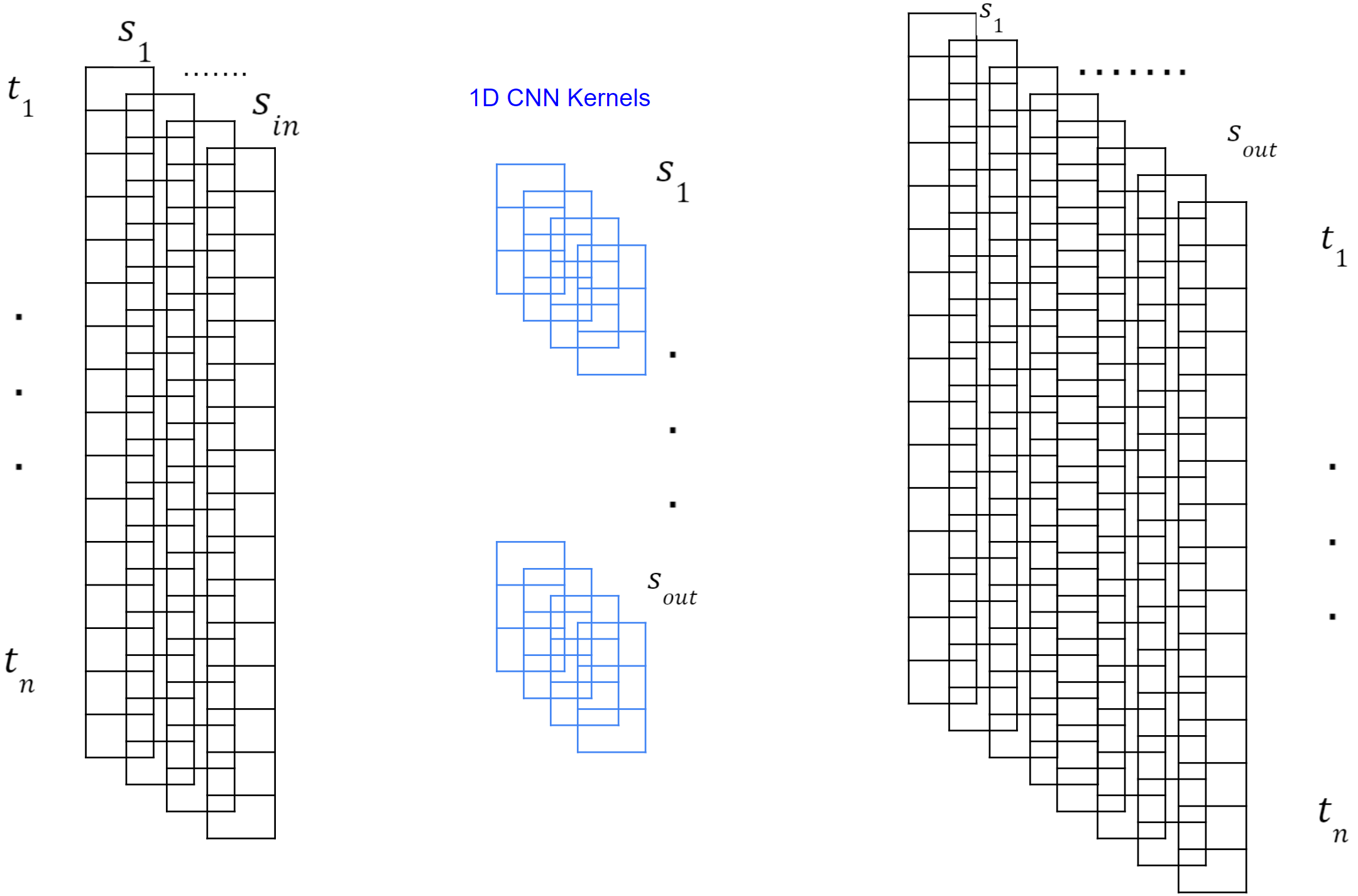}
    \caption{The sequence length stays as \textit{n}, but more sequence patterns have been extracted in each output channel;}
    \label{fig:p3}
\end{figure}

\subsection{Encoder Design Choices}
When forecasting the stock price, the task can be modeled as long sequence time-series forecasting (LSTF) question. The challenges for LSTF include capturing the long-range dependency and efficient operations on capturing the dependencies on long sequence. We consider two choices when designing the encoder.
\begin{figure}[h!]
    \centering
    \includegraphics[scale=0.25, width=50mm]{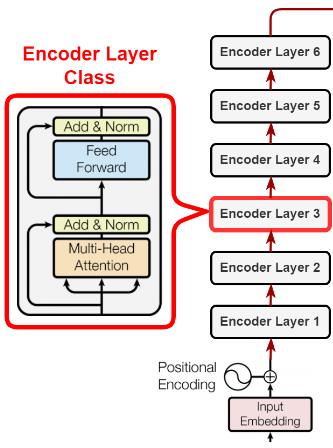}
    \caption{Encoder layers stack together;}
    \label{fig:p4}
\end{figure}
\subsubsection{Full Attention}
Full attention mechanism is applied on the naive Transformer. The length for capturing a dependency on a sequence is theoretically \textit{$O(1)$} which avoids the recurrent structure and outperforms RNN models. However, as shown in Fig \ref{fig:p4}, when numerous encoder layers stack together and each of the attention layer contains a multi-head attention block, the memory usage becomes a bottleneck. Assuming the sequence length is \textit{$L$}, each multi-head attention block will require \textit{$O(L_{}^{2})$} memory space. And if there are J encoder layers stacking together, the memory complexity will be \textit{$O(J * L_{}^{2})$}. This creates higher hardware requirement during the training and makes real-time prediction expensive.
\subsubsection{ProbSparse Attention \& Self-attention Distilling}
Aiming to solve time and memory complexity issues in the naive Transformer. This project considers of using the ProbSparse Attention and Self-attention Distilling techniques from the \textit{Informer} \cite{zhou2021informer}. When calculating the attention score in each multi-head attention layer, a subset number of keys will be selected and follow the attention score equation below:
\begin{figure}[h!]
    \centering
    \includegraphics[scale=0.25, width=50mm]{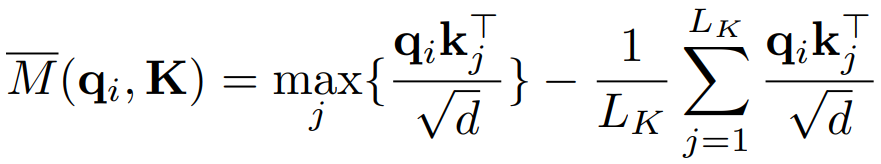}
    \label{fig:p5}
\end{figure}

\noindent As shown in the equation, according to \cite{zhou2021informer}, the top subset number of attention scores will be subtracted by the average attention score across all the queries and selected subset of keys. This will decrease the time and space complexity to \textit{$O(L * log(L))$}. For Self-attention Distilling, as shown in the equation below:
\begin{figure}[h!]
    \centering
    \includegraphics[scale=0.5, width=50mm]{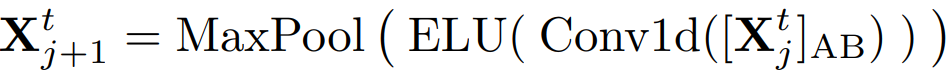}
    \label{fig:p6}
\end{figure}

\noindent At the end of each encoder layer, a max-pooling layer with stride of 2 is added to down-sample the output by half. According to \cite{zhou2021informer}, the total memory usage for the whole encoder structure will be reduced to \textit{$O((2-\varepsilon)Llog(L))$} where $\varepsilon$ is a very small number.

\section{Training Design}
When it comes to train the model, different design choices of loss functions and learning rate schedulers affect the performance of the model in the real world. 

\subsection{Loss Function}
So with stock market prediction the obvious goal is to make money. It is important for our loss function to resemble this goal through an easy to calculate and differentiable function with sufficiently strong gradients. 

For generic time series tasks the goal is the make the model's output predict the target for the next time step. For financial applications knowing the exact price could be important when trading options and other advance financial instruments. Two loss function we tried for this interpretation of the problem are:

\subsubsection{Mean Squared Error} MSE loss is always the first choice in the numerical related task. However, one potential problem with using the MSE is that it can be sensitive to outlier values in the dataset. This means that a single extreme value can have a disproportionately large impact on the overall error. As a result, the MSE may not be the best choice for the hourly stock price since tremendous change rarely happen within one hour.

\subsubsection{Mean Absolute Error} MAE, on the other hand, is very useful in the case where there are a few very large errors and many smaller ones. But, it is not differentiable at 0. In addition, to make the model profitable in the real world, the loss function design need to consider the way of trading the stock and the cumulative profit in the long run.\\

Stepping back though, we realized for our purposes we would really only need to know the direction of stock's movement (in simple terms: is the price going to go up or down). It would also be helpful to have some notion of confidence/magnitude for determining how much of your portfolio you should buy/short the target asset. We created 2 types of logit based trading algorithms. Note that when being used as a loss it is negated to make lower be better.
\subsubsection{Stock Direction} treats the sign of the output of the model as the direction of price movement. This algorithm will simply buy if the sign is positive or short if the sign is negative. There is an optional parameter called threshold where the absolute value of the output has the be above the threshold for us to buy/short. This extremely loosely makes the magnitude of the output resemble confidence. The idea behind the threshold is that even if we know the direction of price movement it is not always good to participate as the unexpected costs and commission fees could make the trade not profitable. The ROI can be calculated as
\begin{equation*}
    ROI = \frac{return}{investment} - 1
\end{equation*}
where
\begin{equation*}
    \frac{return}{investment} = \prod_t{1+sign(output_t)\times PercentChange_t}
\end{equation*}
if using PercentChange or 
\begin{equation*}
    \frac{return}{investment} = exp(\sum_t{sign(output_t)\times LogPercentChange_t})
\end{equation*}
if using LogPercentChange

\subsubsection{Stock Tanh} is the same as \textit{Stock Direction} except that instead of going "all in" it will choose a percent of the portfolio to invest. It chooses this partial investment by processing the model's outputs with the tanh function to be between -1 and 1.
\begin{equation*}
    \frac{return}{investment} = \prod_t{1+tanh(output_t)\times PercentChange_t}
\end{equation*}
or
\begin{equation*}
    \frac{return}{investment} = exp(\sum_t{tanh(output_t)\times LogPercentChange_t})
\end{equation*}

\textit{}

\subsection{Learning Rate Scheduler}
We implemented and experimented with three types of learning rate schedulers to avoid the instability of the model during the training.

\subsubsection{Handcrafted Learning Rate Scheduler} The learning rate will be decreased to a set of empirically selected number when the model has been trained for certain epochs.

\subsubsection{Multiplicative Learning Rate Scheduler} This learning rate scheduler borrows the implementation of MultiplicativeLR from PyTorch. It multiplies the current learning rate by the specified factor at each step. This can help the model converge to a better solution by adjusting the learning rate in response to the duration of the training process.

\subsubsection{Reduce Learning Rate On Plateau} This learning rate scheduler borrows the implementation of ReduceLROnPlateau from PyTorch. It monitors the validation loss and reduces the learning rate when the loss stops improving by the specified amount for the specified number of epochs. This can prevent the model from overfitting to the training data and help to improve the performance of the model on the validation set and ultimately lead to better results on unseen data.

\section{Evaluation Metrics}
This section heavily references the loss section. The way we evaluate the model differs based on the loss method we choose. If we use MSE or MAE we look at the respective aggregate on the whole prediction set. To see if the results are meaningful we can simply compare to if the model just outputted zero every time. We could also apply the stock direction algorithm to evaluate our $\frac{return}{investment}$. If we used the stock direction or stock tanh metric as our loss, we can just use that to figure out our $\frac{return}{investment}$.

\section{Experiment}
In this section, we will discuss and analysis the result we found via manual hyperparameter tuning because of limited time and computing power. In the end, our Stockformer will compare against the zeros and LSTM.

\subsection{Training Phenomenon}

We found that Transformers are harder to train than we originally expected. The choice of learning rate and how it is scheduled seemed to matter way more than we expected. This complicated our experimentation because learning rates do not always transfer between our loss functions. If we choose a learning rate too large the model will gridlock due to the gradients diminishing. We found the cause of the deadlock by monitoring the L2 norm of the gradients to see if it is going to zero or even infinity. We've noticed that the model tends to get stuck in a local optima if the learning rate starts too low. 


\begin{figure}[h!]
    \centering
    \includegraphics[width=\linewidth]{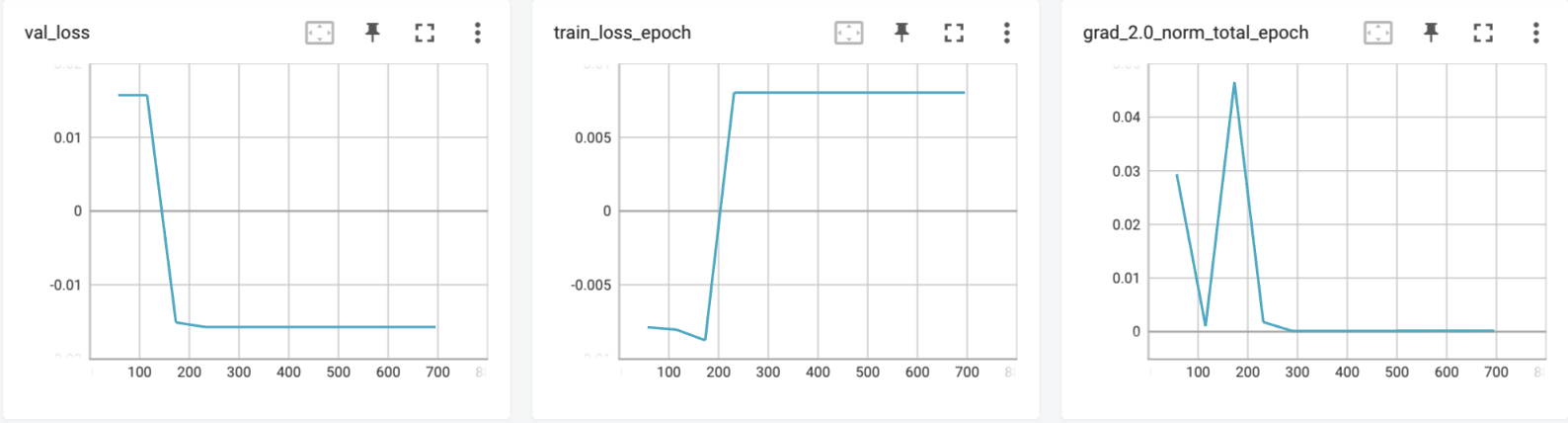}
    \caption{What it looks like when the learning rate is too high (LR=$10^{-4}$); The last chart is of the L2 norm of the gradient; It went to zero and nothing will change;}
    \label{fig:p5}
\end{figure}

\begin{figure}[h!]
    \centering
    \includegraphics[width=\linewidth]{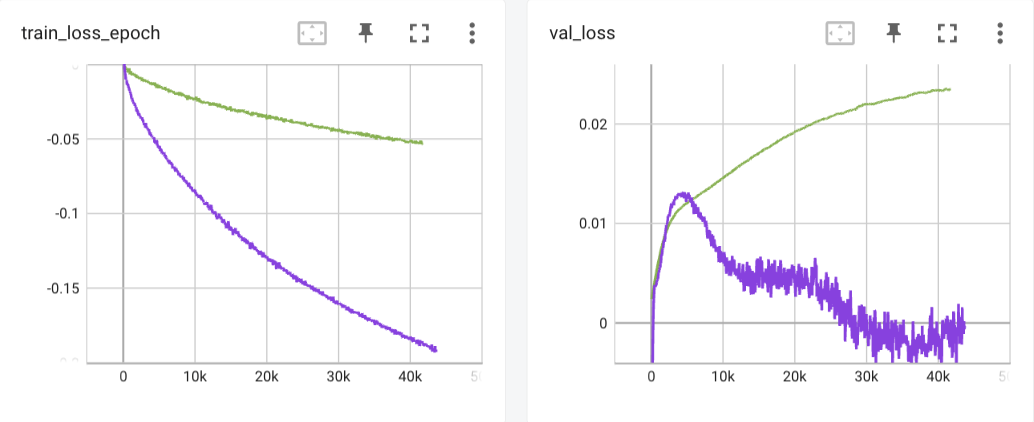}
    \caption{The green data is from Stock Tanh on log percent change with a low LR=$10^{-7}$ while the purple has a decent LR=$10^{-6}$. The validation loss on the purple initially goes up and eventually comes back down and plateaus. The green data seems to be heading into a local optima.}
    \label{fig:p6}
\end{figure}

\noindent On the other hand, if we choose a "decent" learning rate we observe a phenomenon where the training loss decreases as the model start to fit the training data but the validation loss goes up in the beginning forming a plateau and eventually starts to go down and coverages very slowly as the purple line shown in \ref{fig:p6}.

\subsection{Possible Solution}
As we can see, with this "decent" learning rate, the validation loss after the first epoch tends to be the best. Theorically speaking, this can be a sign of skipping the optimal path due to the wrong learning rate. Therefore, it leads us to try to use a really small learning rate or a scheduler that adjusts the learning rate based on the validation loss to overcome this issue. 


\begin{figure}[h!]
    \centering
    \includegraphics[width=\linewidth]{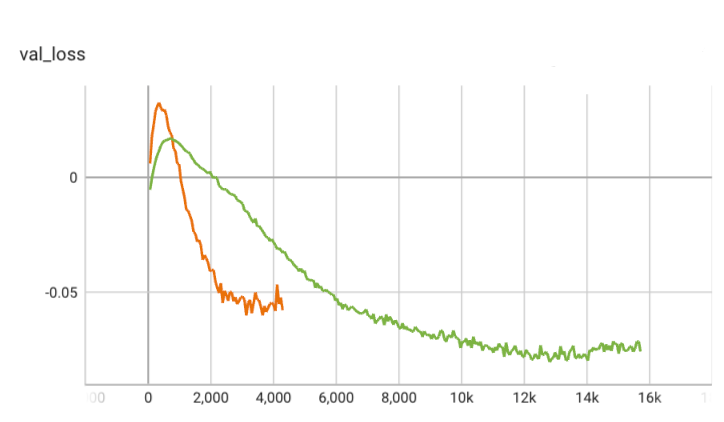}
    \label{fig:p11}
\end{figure}

\noindent The green curve in the figure above shows the best learning rate scheduler (Reduce Learning Rate On Platea) we tried. We can see that the green curve performs better, as it adjusts the learning rate based on validation loss. and the huge plateau seems less significant. However, the curve with the learning rate scheduler still converges slower. Therefore, we start focusing on the other Transformer parameters. In the following section, by tuning the embedding size and the number of attention heads, we are able to solve the plateau issue and converge faster.

\subsection{Table of Hyperparameter vs Profit}
Due to the limited computing power, we were only able to perform the following experiment with these parameter pairs on embedding size and number of attention heads.

\begin{table}[!ht]
    \centering
    \caption{The percent profit is obtained via the Tanh algorithm}
    \begin{tabular}{|l|l|l|l|l|l|l|l|}
    \hline
        (E Size, \# Head) & (128,128) & (256,256) & (512,512) \\ \hline
        pct\_profit & ~ 1.2414 & 1.4788 & ~ 1.7550 \\ \hline
    \end{tabular}
    
\end{table}

\begin{figure}[h!]
    \centering
    \includegraphics[width=\linewidth]{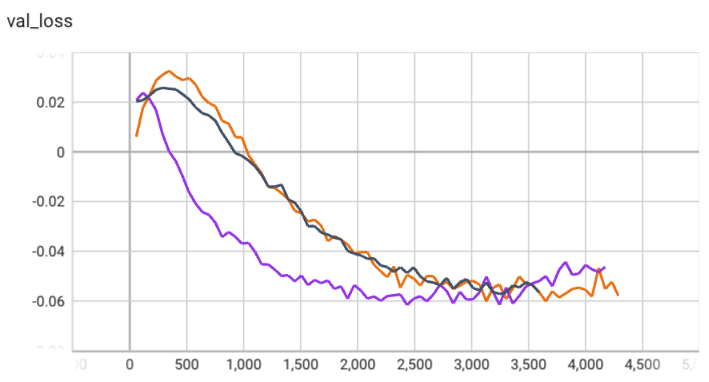}
    \caption{Purple, black, orange, curve:512, 256, 128 embedding size and number of attention head; We can see that the black curve converges fastest;}
    \label{fig:p12}
\end{figure}

\noindent According to our observation in Fig \ref{fig:p12} and assumption, there is a strong relationship between the embedding size and the number of attention heads. The embedding size ideally means the number of time series patterns extracted by the embedding layer. Then, the attention heads will be looking for the patterns among these extracted time series patterns. Therefore, to increase the performance of the model, as the embedding size increases the number of attention heads should also increase.

\subsection{Full Attention and ProbSparse Attention comparisons}
For ProbSparse Attention, when compared against zeros (not executing any trading strategies), the ProbSparse Attention wins because it gains money in both the validation and testing set over the long run. This indicates that the ProbSparse Attention from the Informer generalizes well on predicting the stock even though some insignificant information was ignored during the attention score calculation and the self-attention distilling process. On the other hand, the Full Attention setup has almost the same performance on profit making. Meanwhile, the ProbSparse Attention and the self-attention distilling process can bring better time and space complexity. Therefore, switching to ProbSparse Attention setup is a better choice.

\begin{figure}[h!]
    \centering
    \includegraphics[width=\linewidth]{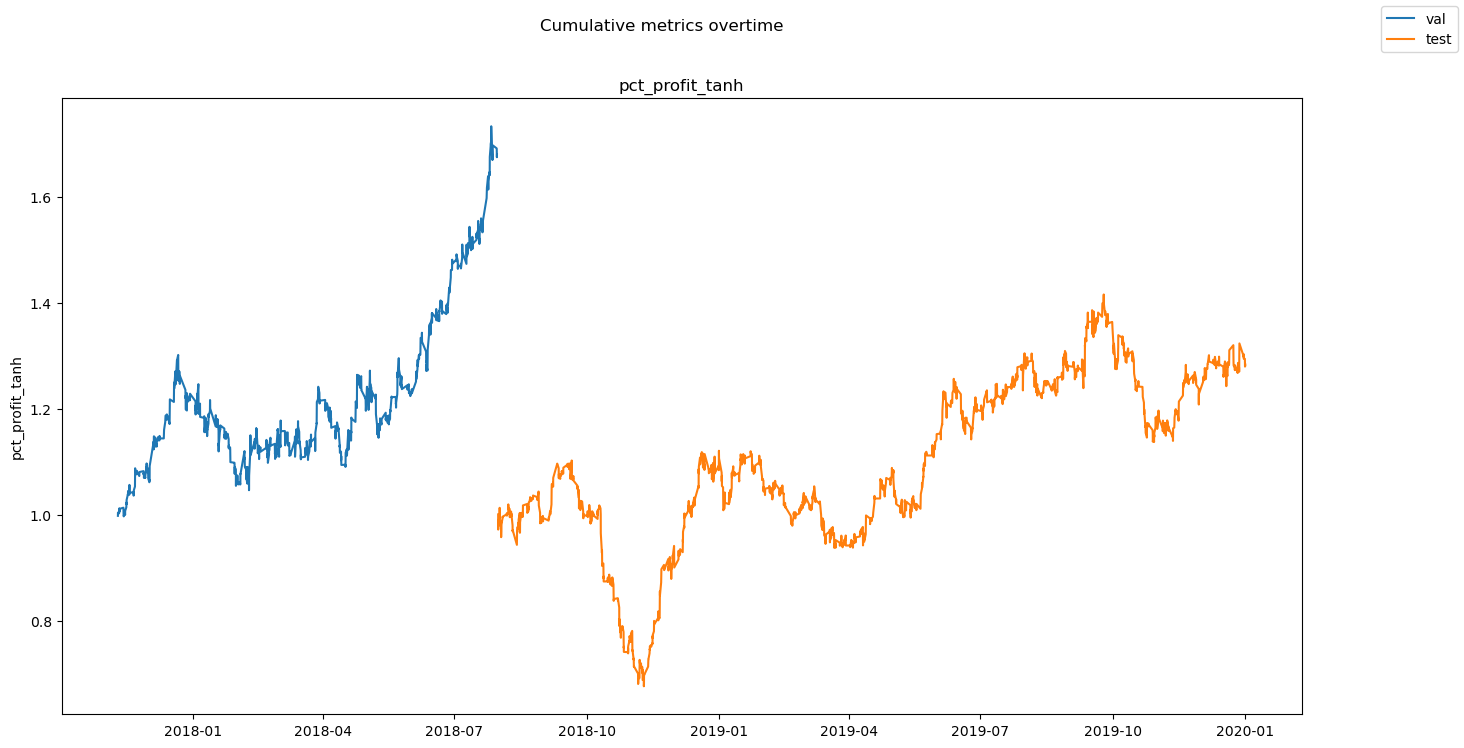}
    \caption{Cumulative Percent of Profit for Stockformer}
    \label{fig:p7}
\end{figure}


\subsection{LSTM comparisons}
In order to show the benefit of using Stockformer in the real world, we compare it with the traditional LSTM model.

\begin{figure}[h!]
    \centering
    \includegraphics[width=\linewidth]{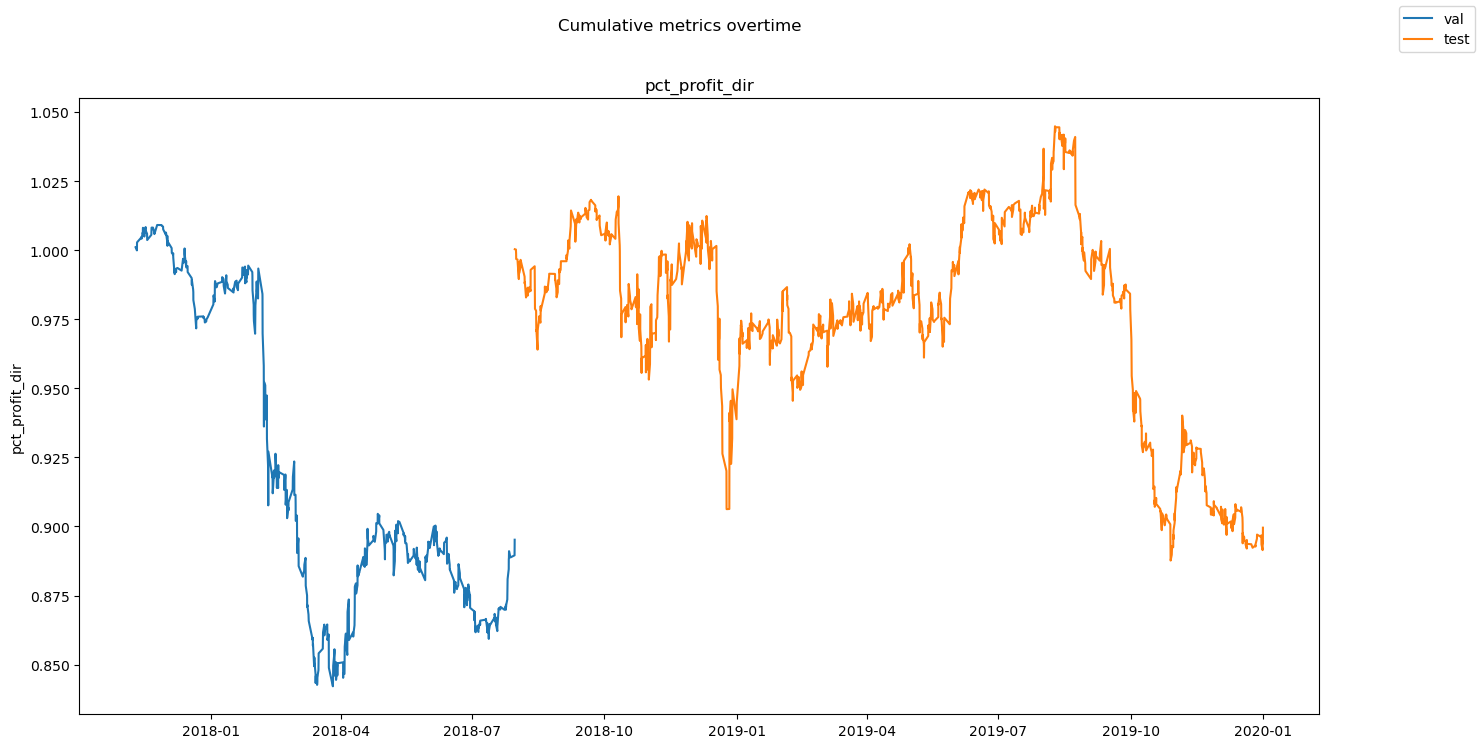}
    \caption{Cumulative Percent of Profit for LSTM}
    \label{fig:p9}
\end{figure}


\noindent As shown in the Fig \ref{fig:p7} and \ref{fig:p9}, overall the Stockformer outperform the LSTM on profit making.


\section{Conclusion \& Future Direction} 
This project is still in its early stage. We are still discovering and implementing more features to it. Although, currently, the model does not guarantee to generate a profitable result, we have found some potential directions to work toward making the model profitable in the future.

\subsubsection{More Tickers} The current input to the model only includes the percent of change of stocks in the oil industry. To show the power of the Stockformer on finding dependencies, more financial tickers such as S\&P 500 Energy, retail oil prices, and AMEX Oil Index should be included. So, the attention mechanism can use their relations to make a better prediction on the target financial ticker. It would also be interesting to look at market sentiment and other alternative indicators.

\subsubsection{Dynamic Training} According to the Efficient Market Hypothesis, we should not expect our model to perform well in 2030 if it is trained on the data from 2010. Therefore, in order to capture the latest pattern on financial tickers, the model should always be retrained with the latest data after a certain period of time. We would also like to back-test the learning algorithm. This is where we would start with some data, run our training algorithm to predict the month following the end of the data, record profit or loss, include the real data for the month that we just predicted into our training data mix, and then repeat this process on the following month. The whole process will end once we get to present day.

\subsubsection{Temporal Encoding}
For our current work we did not use any temporal encoding but instead opted to just use a positional encoding. We strongly believe that including some form of temporal encoding will improve the models output. There are several popular ways to perform the temporal encoding, the most interesting to us being Time2Vec \cite{kazemi2019time2vec}. However as opposed to being just time stamped data each of our data points is a percent change over a timeframe. We'd like to experiment with creating a TimeFrame2Vec. This could lead to more extensions like providing the model with data from multiple time frames, for example, daily or hourly.



\bibliographystyle{plainnat}
\bibliography{references}

@article{brock1992simple,
  title={Simple technical trading rules and the stochastic properties of stock returns},
  author={Brock, William and Lakonishok, Josef and LeBaron, Blake},
  journal={The Journal of finance},
  volume={47},
  number={5},
  pages={1731--1764},
  year={1992},
  publisher={Wiley Online Library}
}

@article{barr1998persuasive,
  title={Persuasive evidence of market inefficiency},
  author={Barr Rosenberg, Kenneth Reid and Lanstein, Ronald},
  journal={Streetwise: the Best of the Journal of Portfolio Management},
  volume={48},
  year={1998},
  publisher={Princeton University Press}
}

@article{KOHZADI1996169,
title = {A comparison of artificial neural network and time series models for forecasting commodity prices},
journal = {Neurocomputing},
volume = {10},
number = {2},
pages = {169-181},
year = {1996},
note = {Financial Applications, Part I},
issn = {0925-2312},
doi = {https://doi.org/10.1016/0925-2312(95)00020-8},
url = {https://www.sciencedirect.com/science/article/pii/0925231295000208},
author = {Nowrouz Kohzadi and Milton S. Boyd and Bahman Kermanshahi and Iebeling Kaastra}
}

@inproceedings{zhou2021informer,
  title={Informer: Beyond efficient transformer for long sequence time-series forecasting},
  author={Zhou, Haoyi and Zhang, Shanghang and Peng, Jieqi and Zhang, Shuai and Li, Jianxin and Xiong, Hui and Zhang, Wancai},
  booktitle={Proceedings of the AAAI Conference on Artificial Intelligence},
  volume={35},
  number={12},
  pages={11106--11115},
  year={2021}
}

@article{as2013study,
  title={A study on fundamental and technical analysis},
  author={AS, Suresh},
  journal={International Journal of Marketing, Financial Services \& Management Research},
  volume={2},
  number={5},
  pages={44--59},
  year={2013}
}

@article{501826,
  author={Schierholt, K. and Dagli, C.H.},
  booktitle={IEEE/IAFE 1996 Conference on Computational Intelligence for Financial Engineering (CIFEr)}, 
  title={Stock market prediction using different neural network classification architectures}, 
  year={1996},
  volume={},
  number={},
  pages={72-78},
  doi={10.1109/CIFER.1996.501826}
}

@article{8488440,
  author={Namdari, Alireza and Li, Zhaojun Steven},
  booktitle={2018 IEEE Technology and Engineering Management Conference (TEMSCON)}, 
  title={Integrating Fundamental and Technical Analysis of Stock Market through Multi-layer Perceptron}, 
  year={2018},
  volume={},
  number={},
  pages={1-6},
  doi={10.1109/TEMSCON.2018.8488440}
}

@article{ismail2019deep,
  title={Deep learning for time series classification: a review},
  author={Ismail Fawaz, Hassan and Forestier, Germain and Weber, Jonathan and Idoumghar, Lhassane and Muller, Pierre-Alain},
  journal={Data mining and knowledge discovery},
  volume={33},
  number={4},
  pages={917--963},
  year={2019},
  publisher={Springer}
}

@inproceedings{wang2017time,
  title={Time series classification from scratch with deep neural networks: A strong baseline},
  author={Wang, Zhiguang and Yan, Weizhong and Oates, Tim},
  booktitle={2017 International joint conference on neural networks (IJCNN)},
  pages={1578--1585},
  year={2017},
  organization={IEEE}
}

@ARTICLE{726791,
  author={Lecun, Y. and Bottou, L. and Bengio, Y. and Haffner, P.},
  journal={Proceedings of the IEEE}, 
  title={Gradient-based learning applied to document recognition}, 
  year={1998},
  volume={86},
  number={11},
  pages={2278-2324},
  doi={10.1109/5.726791}
}

@article{chung2014empirical,
  title={Empirical evaluation of gated recurrent neural networks on sequence modeling},
  author={Chung, Junyoung and Gulcehre, Caglar and Cho, KyungHyun and Bengio, Yoshua},
  journal={arXiv preprint arXiv:1412.3555},
  year={2014}
}

@article{hochreiter1997long,
  title={Long short-term memory},
  author={Hochreiter, Sepp and Schmidhuber, J{\"u}rgen},
  journal={Neural computation},
  volume={9},
  number={8},
  pages={1735--1780},
  year={1997},
  publisher={MIT Press}
}

@article{HOSEINZADE2019273,
title = {CNNpred: CNN-based stock market prediction using a diverse set of variables},
journal = {Expert Systems with Applications},
volume = {129},
pages = {273-285},
year = {2019},
issn = {0957-4174},
doi = {https://doi.org/10.1016/j.eswa.2019.03.029},
url = {https://www.sciencedirect.com/science/article/pii/S0957417419301915},
author = {Ehsan Hoseinzade and Saman Haratizadeh}
}

@article{gjylapirecurrent,
  title={Recurrent Neural Networks in Time Series Prediction},
  author={Gjylapi, Dezdemona and Proko, Eljona and Hyso, Alketa}
}

@article{lu2020cnn,
  title={A CNN-LSTM-based model to forecast stock prices},
  author={Lu, Wenjie and Li, Jiazheng and Li, Yifan and Sun, Aijun and Wang, Jingyang},
  journal={Complexity},
  volume={2020},
  year={2020},
  publisher={Hindawi}
}

@article{vaswani2017attention,
  title={Attention is all you need},
  author={Vaswani, Ashish and Shazeer, Noam and Parmar, Niki and Uszkoreit, Jakob and Jones, Llion and Gomez, Aidan N and Kaiser, {\L}ukasz and Polosukhin, Illia},
  journal={Advances in neural information processing systems},
  volume={30},
  year={2017}
}

@misc{hochreiter2001gradient,
  title={Gradient flow in recurrent nets: the difficulty of learning long-term dependencies},
  author={Hochreiter, Sepp and Bengio, Yoshua and Frasconi, Paolo and Schmidhuber, J{\"u}rgen and others},
  year={2001},
  publisher={A field guide to dynamical recurrent neural networks. IEEE Press In}
}

@article{diks2006new,
  title={A new statistic and practical guidelines for nonparametric Granger causality testing},
  author={Diks, Cees and Panchenko, Valentyn},
  journal={Journal of Economic Dynamics and Control},
  volume={30},
  number={9-10},
  pages={1647--1669},
  year={2006},
  publisher={Elsevier}
}

@article{kazemi2019time2vec,
  title={Time2vec: Learning a vector representation of time},
  author={Kazemi, Seyed Mehran and Goel, Rishab and Eghbali, Sepehr and Ramanan, Janahan and Sahota, Jaspreet and Thakur, Sanjay and Wu, Stella and Smyth, Cathal and Poupart, Pascal and Brubaker, Marcus},
  journal={arXiv preprint arXiv:1907.05321},
  year={2019}
}

\end{document}